\documentclass[letterpaper, 10 pt, conference]{ieeeconf}
\IEEEoverridecommandlockouts
\usepackage{amsfonts}
\usepackage{amsmath}
\usepackage{graphicx}
\usepackage{subcaption}
\usepackage[ruled]{algorithm2e}
\usepackage{verbatim}
\usepackage{amssymb}
\usepackage{blkarray, bigstrut}

\newtheorem{theorem}{Theorem}
\newtheorem{prop}{Proposition}
\newtheorem{definition}{Definition}

\DeclareMathOperator*{\argmax}{arg\,max}

\newcommand{\opt}{{\rm opt}}

\newcommand{\eq}{{\rm eq}}
\newcommand{\nei}{{\mathcal{N}_i}}
\newcommand{\mc}{{\rm MC}}

\newcommand{\poa}{\ensuremath{{\rm PoA}}}
\newcommand{\nee}{{\cal N}}

\title{Valid Utility Games with Information Sharing Constraints}
\author{David Grimsman, Philip N. Brown, and Jason R. Marden %
\thanks{D. Grimsman (\texttt{grimsman@cs.byu.edu}) is with the Computer Science Department at Brigham Young University, Provo, UT}
\thanks{P. N. Brown (\texttt{philip.brown@uccs.edu}) is with the Department of Computer Science, University of Colorado, Colorado Springs, CO}
\thanks{J. R. Marden (\texttt{jmarden@ece.ucsb.edu}) is with the Department of Electrical and Computer Engineering, University of California, Santa Barbara, CA}
\thanks{© 2022 IEEE. Personal use of this material is permitted. Permission from IEEE must be
obtained for all other uses, in any current or future media, including
reprinting/republishing this material for advertising or promotional purposes, creating new
collective works, for resale or redistribution to servers or lists, or reuse of any copyrighted
component of this work in other works.}
}

\begin{document}
	
\maketitle

\begin{abstract}
    The use of game theoretic methods for control in multiagent systems has been an important topic in recent research. Valid utility games in particular have been used to model real-world problems; such games have the convenient property that the value of any decision set which is a Nash equilibrium of the game is guaranteed to be within 1/2 of the value of the optimal decision set. However, an implicit assumption in this guarantee is that each agent is aware of the decisions of all other agents. In this work, we first describe how this guarantee degrades as agents are only aware of a subset of the decisions of other agents. We then show that this loss can be mitigated by restriction to a relevant subclass of games.
\end{abstract}

\section{Introduction}

Game theoretic methods have recently received much attention for their usefulness in the control of multiagent systems \cite{Marden2015a}. These methods have been applied in a wide variety of applications, including air traffic control \cite{Castelli2011,Vossen2002}, the allocation of resources at universities \cite{Ghosh2013}, sensor placement \cite{Krause2008}, airport security \cite{Pita2008}, the power grid \cite{Saad2012}, and radio networks \cite{Xu2012}. A common paradigm is that each agent is a player in a cooperative game. Each agent's decision-making is governed by a utility function implemented by the system designer. The goal is to create utility functions that maximize a social welfare function valued by the system designer.

An important class of models is that of valid utility games \cite{Vetta2002,Bachrach2012EfficiencyWelfare,Chen2011TheGames}. Such games require that (1) the social welfare function exhibits \emph{submodularity}, a certain ``diminishing returns" property, (2) an agent's utility is at least its marginal contribution to the social welfare, and (3) the total utility is at most the total value of social welfare. In this work, we also include the property that the welfare function is nondecreasing. Given these restrictions, the model is still widely applicable and has been used to analyze market sharing \cite{Goemans2005}, allocating scientific credit \cite{Kleinberg2011MechanismsCredit}, facility location 
\cite{Vetta2002}, and network formation games \cite{2007AlgorithmicTheory}, among others.

A key result in valid utility games is that the value (according to the welfare function) of any Nash equilibrium set of agent decisions is guaranteed to be within 1/2 of the value of the optimal decision set \cite{Vetta2002}. In other words, the system designer may choose any set of utility functions and welfare function that satisfy the given properties, and the emergent behavior has this optimality guarantee, regardless of the number of agents in the system.

An underlying assumption in this guarantee, however, is that each agent knows the decisions of all other agents in the system. In many real-world scenarios, this assumption may be unrealistic. For instance, communication could be constrained by limited bandwidth \cite{Grimsman2019TheMaximization}, the number of agents could be too large \cite{mirzasoleiman2016}, information could be contaminated or restricted by an attacker \cite{Grimsman2020TheOptimization}, there could be a lack of trust among the agents \cite{Shokri2016a}, or the allowable amount of time could be limited \cite{Sun2020DistributedExecution}. Therefore, recent work has begun to explore what happens in settings where each agent is only aware of a subset of other agents. For instance \cite{Grimsman2020TheOptimization} considers what happens to the $1/2$ guarantee in valid utility games when a subset of agents is compromised by an attacker, i.e., they can no longer communicate with the group, or vice versa. The paper shows that with each compromised agent, the denominator in the guarantee increases by $1$, e.g., for one compromised agent the guarantee decreases to $1/3$. The work in \cite{Grimsman2019TheMaximization} considers a slightly different setting: that of the greedy submodular maximization, wherein the agents have an implicit ordering and make decisions sequntially according to that ordering. The information sharing constraints are more broad, however, in that each agent can only base its decision on a subset of agents previous in the sequence. It was shown that the guarantees of the algorithm decreased as the maximum number of agents among whom there was no information sharing increased.

This work generalizes these two works: here we consider the general class of all valid utility games and also allow for broad information sharing constraints, i.e., each agent can observe the actions of some subset of agents. Two natural questions arise:
\begin{enumerate}
	\item How does the structure of these information sharing constraints affect the equilibrium performance guarantees?
	\item What utility functions can provide a better guarantee?
\end{enumerate}

Theorem~\ref{thm:vug_poa} addresses the first question: essentially, the performance guarantee is inversely related to the number of groups of agents who have access to the same information -- for instance, in the complete graph, this number is 1. This number increases with the number of constraints on information sharing, and we show for a system with $n$ agents, a strategic attacker could cause (with minimal intervention) the performance guarantee to fall to $1/(n+1)$, arbitrarily bad for large systems.

Theorem~\ref{thm:cons} addresses the second question by introducing a subset of valid utility games, where the utility functions must all satisfy a certain consistency property. Among this smaller set of systems, a lower bound on the guarantee is (similar to \cite{Grimsman2019TheMaximization}) instead tied to the largest set of agents, among whom there is no reciprocal information sharing. This lower bound is at least as high as the guarantee in Theorem~\ref{thm:vug_poa}, and is strictly better for most problem instances. Additionally, Proposition~\ref{prop:svug_ub} provides an upper bound on any possible guarantee for any fixed set of utility functions, even those outside the bounds of valid utility games. This bound is also given in terms of the largest set of agents among whom there are not any edges.

\section{Model} \label{sec:model}
Consider a set of base elements $S$ and a set of agents $N = \{1, \dots, n\}$. Agent $i$ has access to a set of actions $X_i$, where each action $x_i \in X_i$ is a subset of elements of $S$. Additionally, agent $i$ is endowed with a utility function $U_i:X_1 \times \cdots \times X_n \to \mathbb{R}$, which evaluates the action $x_i$, dependent on the decisions of all other agents, which we will denote as $x_{-i} := (x_1, \dots, x_{i-1}, x_{i+1}, \dots, x_n)$. To highlight this dependence, we use the notation $U_i(x_i, x_{-i})$.

An action profile $x=(x_1, \dots, x_n)$ is evaluated by a welfare function $w:X_1 \times \cdots \times X_n \to \mathbb{R}$. Thus the agents' collective goal is to find action profile $x^\opt$ such that
\begin{equation}
    x^\opt \in \argmax_{x_1 \in X_1, \dots, x_n \in X_n} w(x_1, \dots, x_n)
\end{equation}
In this work we consider welfare functions of the form $w(x) = f\left( \cup_i x_i\right)$, where $f:2^S \to \mathbb{R}$. We will abuse notation and denote $f(x_i, x_j)$ to mean $f(x_i \cup x_j)$. Our focus will be on \emph{valid utility games} \cite{Vetta2002}:

\begin{definition} \label{def:vug}
	A Valid Utility Game (VUG) is a system with no information sharing constraints that satisfies the following three conditions:
	\begin{enumerate}
	    \item $f$ is\footnote{The original defintion used in \cite{Vetta2002} did not require monotonicity or normalization, yet we impose that as it suits our purposes here.} \label{itm:submod}
	    \begin{enumerate}
	        \item \emph{submodular}: \\ $f\left(A \cup \{s\}\right) - f(A) \ge f\left(B \cup \{s\}\right) - f(B)$ for all $A \subseteq B \subseteq S$ and $s \in S \setminus B$.
            \item \emph{nondecreasing}: $f(A) \le f(B)$ for all $A \subseteq B \subseteq S$
            \item \emph{normalized}: $f(\emptyset) = 0$.
	    \end{enumerate}
	    \item $U_i(x_i, x_{-i})\geq f(x_i, x_{-i})-f\left(x_{-i}\right)$ for all $i$\label{itm:marg}
	    \item $\sum_i U_i(x_i, x_{-i})\leq f(x_i, x_{-i})$ \label{itm:sum}
	\end{enumerate}
\end{definition}
As an example, consider the marginal contribution utility, defined as:
\begin{equation} \label{eq:mcdef}
    U_i(x_i, x_{-i}) = f(x_i, x_{-i}) - f(x_{-i}).
\end{equation}
If each agent implements this utility, obviously property \ref{itm:marg}) is satisfied at equality, and one can show that the submodularity of $f$ implies that \ref{itm:sum}) is likewise satisfied.

The emergent behavior of the system is defined to be a Nash equilibrium. An action profile $x^\eq$ is a Nash equilibrium if
\begin{equation} \label{eq:eqdef}
    U_i(x_i^\eq, x_{-i}^\eq) \ge U_i(x_i, x_{-i}^\eq), \text{ for all } x_i \in X_i \text{ and all } i.
\end{equation}
In some contexts, the above definition is called a \emph{pure strategy Nash equilibrium}, while a probability distribution across all action profiles $x$ is called a mixed strategy, with an analogous definition of a \emph{mixed strategy Nash equilibrium}. A pure strategy Nash equilibrium is not always guaranteed to exist for VUGs, however, a mixed strategy one is. The analysis in this paper focuses on systems that exhibit a pure strategy Nash equilibrium, and we conjecture that this can be extended to mixed strategies -- a subject of future work.

\subsection{Graph Constraints}

The definition of the utility function $U_i(x_i, x_{-i})$ implicitly assumes that each agent has access to the actions of all other agents in the system. In many real-world settings, however, this assumption may not hold. We model this restriction by imposing that agent $i$ is only aware of the actions of some subset of agents $\nei$, and assuming that $x_j = \emptyset$ for $j \in N \setminus \nei$. This assumption can either be thought of as not being aware that these agents exist or that the agents have opted out of the game. We will use the notation $U_i(x_i, x_\nei)$ to highlight this restriction. The definition of the Nash equilibrium can then be extended as follows: an action profile $x^\eq$ is an equilibrium if
\begin{equation} \label{eq:geqdef}
    U_i(x_i^\eq, x_\nei^\eq) \ge U_i(x_i, x_\nei^\eq), \text{ for all } x_i \in X_i \text{ and all } i.
\end{equation}

The sets $\nei$ define a directed graph $G = (V, E)$, where $V = N$ and $(j, i) \in E$ if $j \in \nei$. We refer to such a graph as an \emph{information sharing constraint graph}, as it effectively represents the information sharing constraints on the system. We can now conclude that a game $H$ is fully defined as the tuple $(N, f, \{U_i\}_i, G)$, and we denote $\cal{H}$ to be the set of all games as defined in this section.

Our goal in this work is to understand how an information sharing constraint graph $G$ will affect the value of the resulting equilibria normalized against the value of the optimal solution across all games, i.e., the price of anarchy:
\begin{equation}
    \poa(\mathcal{H}, G) := \min_{\substack{H \in \mathcal{H}(G) \\ x^\eq \in EQ(H)}} \frac{f(x^\eq)}{f(x^\opt(H))} \in [0, 1],
\end{equation}
where $\mathcal{H}(G)$ are the set of all games with information constraint graph $G$ and $EQ(H)$ is the set of all equilibrium profiles that satisfy \eqref{eq:geqdef}. For a graph $G$, a price of anarchy of 1 would mean that all games subject to those information sharing constraints would have optimal emergent behavior, and a price of anarchy close to 0 would mean that the constraints are such that one cannot make any meaningful guarantee about the performance of the emergent behavior. When $G$ is the complete graph, it has been shown that $\poa(\mathcal{H}, G) = 1/2$ \cite{Vetta2002}. Thus we seek to find how various graph structures might further degrade this value.

\section{The Price of Anarchy for Valid Utility Games}

\begin{figure*}
	\centering
	\includegraphics[scale=0.5]{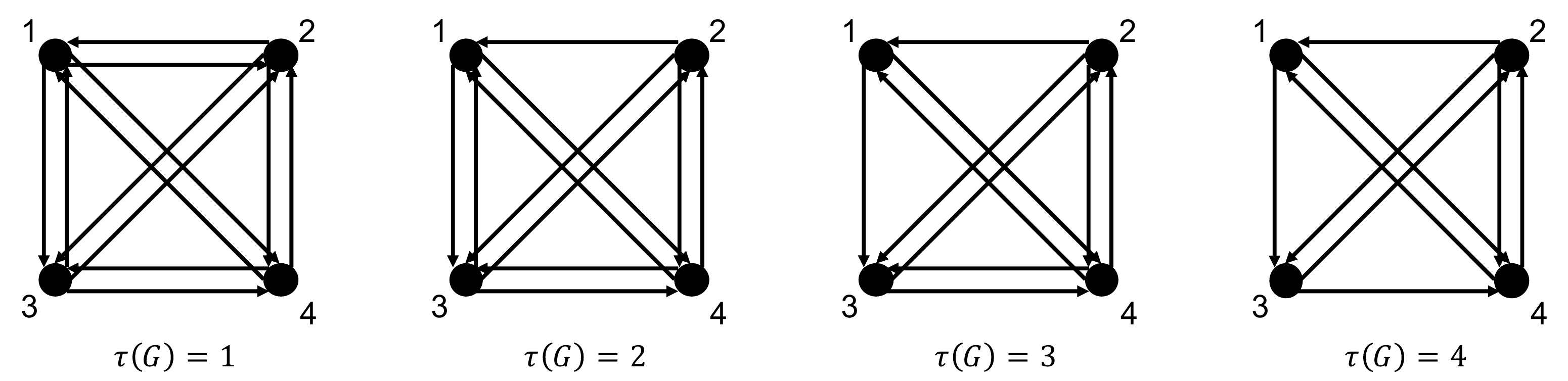}
	\caption{An illustration of information groups. The first graph is a complete graph, thus all nodes are in the only information group. The second graph has edge (1, 2) removed, thus $T(G) = (\{1, 3, 4\}, \{2\})$, since 2 is the only node without an incoming edge from 1. The third graph further has edge (3, 1) removed, and $T(G) = \{3, 4\}, \{1\}, \{2\}$. Finally the last graph has edge (4, 3) removed, and each node is its own information group.}
	\label{fig:info_grp}
\end{figure*}

In this section, we show the price of anarchy as a function of the information constraint graph $G$. To do so, we introduce the notion of an \emph{information group}: a set of nodes which is fully connected, and which have the same incoming neighbors. Formally stated, $T \subseteq N$ is an information group of $G$ if for all $i, j \in T$, $\nei \cup \{i\} = \mathcal{N}_j \cup \{j\}$. An alternate definition is that if $A(G)$ is the adjacency matrix of $G$ and $I$ the identity matrix, then all rows of $A(G) + I$ associated with the nodes in $T$ are the same. A \emph{maximal information group} is an information group that is not a subset of any other information group. We denote the set of all maximal information groups of $G$ as $\mathcal{T}(G)$, which is both unique and a partition on the nodes of $G$. Denote $\tau(G) = |\mathcal{T}(G)|$. See Figure~\ref{fig:info_grp} for an example. The price of anarchy for the set of games subject to a graph constraint $G$ can be expressed in terms of its information group number $\tau$.

\begin{theorem} \label{thm:vug_poa}
	For any $G$ where $G$ has at least one edge,
	\begin{equation} \label{eq:vug_poa}
		\poa(\mathcal{H}, G) = \frac{1}{1 + \tau(G)} \ge \frac{1}{n+1}.
	\end{equation}
\end{theorem}

The full proof is given below, but here we give an overview. The properties from Definition~\ref{def:vug} are used to establish that $\poa(\mathcal{H}, G) \ge \frac{1}{1 + \tau(G)}$. Tightness is shown by example, where one carefully constructs a problem instance and set of utility functions that satisfy Definition~\ref{def:vug} so that agents are incentivized into making poor decisions.

Theorem~\ref{thm:vug_poa} effectively shows that the general class of valid utility games is not robust against these types of information constraints. For instance, consider the example set forth in Figure~\ref{fig:info_grp}. In the leftmost graph, which is a complete graph, we see that $\poa(\mathcal{H}, G) = 1/2$, recovering the well-known result from \cite{Vetta2002}. However, the rightmost graph only has 3 edges removed, yet $\tau(G) = 4$, and $\poa(\mathcal{H}, G) = 1/5$. In fact, for any number of agents this example is instructive: there exist $n-1$ edges that can be removed form the complete graph such that $\poa(\mathcal{H}, G)$ moves from 1/2 to $1/(n+1)$ --- arbitrarily bad. For large systems, this implies that the system designer cannot be content to simply choose utilities that satisfy Definition~\ref{def:vug}.

\begin{proof}
    We first show that
    \begin{equation}
    	\frac{f(x^\eq)}{f(x^\opt)} \ge \frac{1}{1 + \tau(G)},
    \end{equation}
    and then show that for any $G$, there exists $f, \{X_i\}_i, \{U_i\}_i$ which make the expression tight. We denote $\nee_T$ to mean the set of incoming neighbors common to information group $T$. 
    Begin with defining the marginal contribution function $\Delta(A|B): f(A \cup B) - f(B)$, for $A, B \subseteq S$. This give the objective function value for adding the elements in set $A$ to those in set $B$. Then
    \begin{align}
    	f(x^\opt) \le & f(x^\eq) + \Delta(x^\opt | x^\eq)  \label{eq:vug_poa_lb1},\\
    	= & f(x^\eq) + \sum_i \Delta(x^\opt_i | x^\opt_{1:i-1}, x^\eq) \label{eq:vug_poa_lb2},\\
    	\le & f(x^\eq) + \sum_i \Delta(x^\opt_i | x^\eq) \label{eq:vug_poa_lb3}, \\
    	= & f(x^\opt) + \sum_{T \in \mathcal{T}(G)} \sum_{i \in T} \Delta(x^\opt_i | x^\eq) \label{eq:vug_poa_lb4}, \\
    	\le & f(x^\eq) + \sum_{T \in \mathcal{T}(G)} \sum_{i \in T} \Delta(x^\opt_i | x^\eq_{j \in \nee_T}) \label{eq:vug_poa_lb5}, \\
    	\le & f(x^\eq) + \sum_{T \in \mathcal{T}(G)} \sum_{i \in T} U_i(x^\opt_i , x^\eq_{j \in \nee_T}) \label{eq:vug_poa_lb6}, \\
    	\le & f(x^\eq) + \sum_{T \in \mathcal{T}(G)} \sum_{i \in T} U_i(x^\eq_i , x^\eq_{j \in \nee_T}) \label{eq:vug_poa_lb7}, \\			
    	\le & f(x^\eq) + \sum_{T \in \mathcal{T}(G)} f(x^\eq_T) \label{eq:vug_poa_lb8}, \\
    	\le & f(x^\eq)(1 + \tau(G)) \label{eq:vug_poa_lb9},
    \end{align}
    where \eqref{eq:vug_poa_lb1} is true by monotonicity of $f$, \eqref{eq:vug_poa_lb2} is true by definition of $\Delta(\cdot)$, \eqref{eq:vug_poa_lb3} is true by submodularity of $f$, \eqref{eq:vug_poa_lb4} is a reorganization of the sum, \eqref{eq:vug_poa_lb5} is true by submodularity, \eqref{eq:vug_poa_lb6} is true by statement 2) in Definition~\ref{def:vug}, \eqref{eq:vug_poa_lb7} is true by the defintion of equilibrium, \eqref{eq:vug_poa_lb8} is true by statement 3) in the Definition~\ref{def:vug}, and \eqref{eq:vug_poa_lb9} is true by monotonicity of $f$.
    
    \begin{figure}
    	\centering
    	\includegraphics[scale=0.55]{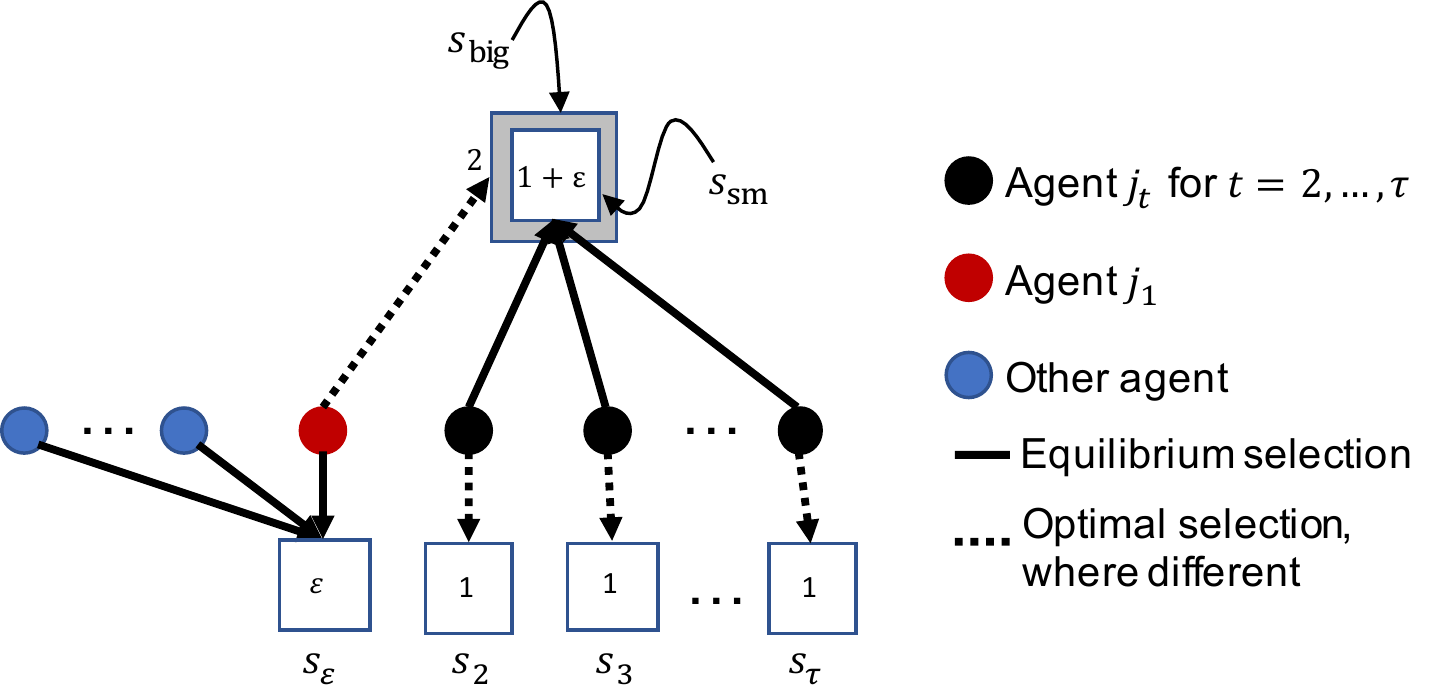}
    	\caption{Example used in the proof for Theorem~\ref{thm:vug_poa}. While agent $j_1$'s choice in equilibrium may seem unintuitive, it is based on a carefully crafted utility function, such that Defintion~\ref{def:vug} is still satisfied for the system.}
    	\label{fig:vug_poa_ub}
    \end{figure}
    
    Next we construct an example worst-case $f$, $\{X_i\}_i$, $\{U_i\}_i$ such that
    \begin{equation} \label{eq:vug_poa_ub}
    	\frac{f(x^\eq)}{f(x^\opt)} = \frac{1}{1 + \tau(G)}
    \end{equation}
    for any $G$. Let $S = \{s_\varepsilon, s_{\rm sm}, s_{\rm big}, s_2, \dots, s_\tau\}$ be a set of (possibly overlapping) 2-D boxes, as shown in Figure~\ref{fig:svug_ub}. Let $f(x)$ be the total area covered by the boxes in $S(x)$: this function is normalized, submodular, and monotone. For some small $\varepsilon > 0$, let $f(\{s_\varepsilon\})  = \varepsilon$, $f(\{s_{\rm sm}\}) = 1+\varepsilon$, $f(\{s_{\rm big}\})  = 2$, and $f(\{s_2\}) =  \cdots = f(\{s_\tau\}) = 1$. The box $s_{\rm big}$ ``covers" the box $s_{\rm sm}$, i.e., $f(\{s_{\rm sm}, s_{\rm big}\}) = 2$. The remaining pairs of boxes are disjoint.
    
    From each information group $T_1, \dots, T_\tau$ choose a representative agent $j_1, \dots, j_\tau$. Since the label order is arbitrary, we assume without loss of generality that there exists incoming edges from agent $j_2$ to the agents in $T_1$,  i.e., $j_2 \in \nee(T_1)$. The action sets are allocated as
    \begin{equation}
    	X_i = \left\{ \begin{array}{ll}
    		\{\{s_\varepsilon\}, \{s_{\rm big}\}\} & \text{if } i=j_1, \\
    		\{\{s_{\rm sm}\}, \{s_t\}\} & \text{if } i = j_t \text{ and } t > 1, \\
    		\{\{s_\varepsilon\}\} & \text{otherwise.}
    	\end{array}
    	\right.
    \end{equation}
    Again, both $f$ and $\{X_i\}_i$ are represented in Figure~\ref{fig:vug_poa_ub}.
    
    In order to define the utilities, we first define the action profile $x^\eq$:
    \begin{equation}
    	x^\eq = \left\{ \begin{array}{ll}
    		\{s_{\rm sm}\} & \text{if } i \in \{j_2, \dots, j_\tau\}, \\
    		\{s_\varepsilon\} & \text{if } i \notin \{j_2, \dots, j_\tau\}. \\
    	\end{array}
    	\right.
    \end{equation}
    As the notation implies, we will design the utilities so that this action profile is an equilibrium. For every $T_t \in \mathcal{T}(G)$ define $x^t$, where
    \begin{equation}
    	x^t_i = \left\{ \begin{array}{ll}
    		x^\eq_i & \text{if } i \in T_t \cup \nee(T_t), \\
    		\emptyset & \text{if } i \notin T_t \cup \nee(T_t).
    	\end{array}
    	\right.
    \end{equation}
    In other words, $x^t$ is the set of actions in $x^\eq$, with the exception that all agents not in $T_t$ choose the empty set. It is important to note that due to the graph constraints, the utility of agent $i \in T_t$ for action profile $x^\eq$ is $U_i(x^t)$. It can also be observed that $f(x^t) = 1 + \varepsilon + p_t\varepsilon$, where $p_t \in \{0, 1\}$ is an indicator: $p_t =1$ if $j_1 \in \nee(T_t)$ or $|T_t|>1$ (i.e., $s_\varepsilon$ is chosen by some agent in $T_t \cup \nee(T_t)$), and $p_t=0$ otherwise.
    
    The utility functions are as follows:
    \begin{equation}
    	U_i(x_i, x_\nei) = \left\{ \begin{array}{ll}
    		1 + \varepsilon & \text{if } x = x^t \text{ and } i = j_t \\
    		& \text{for some } T_t \in \mathcal{T}(G) \\
    		f(x_i, x_{-i}) - f(x_{-i}) & \text{otherwise.}
    	\end{array}
    	\right.
    \end{equation}
    We claim that $f, \{X_i\}, \{U_i\}_i$ is a VUG. Since $U_i = \mc_i$ for all action profiles except when $x = x^t$ and $i=j_t$, we need only prove that the statements 2) and 3) in the VUG definition are satisfied for these exceptions. Statement 2) holds since $f(x^t) - f(x^t_{i \ne j_t}) \le 1 + \varepsilon = U_{j_t}(x^t)$. Statement 3) holds, since:
    \begin{align}
    	\sum_i U_i (x^t) = & U_{j_t}(x^t)  + \sum_{i \notin \{j_t, j_1\}} U_i(x^t) \\
    	\le & 1 + \varepsilon + p_t\varepsilon \\
    	= & f(x^t).
    \end{align}
    
    Recall that agent $j_1$'s action set is $\{\{s_{\rm big}\}, \{s_\varepsilon\}\}$, implying that
    \begin{align}
    	U_{j_1}(x^1) =& 1 + \varepsilon \\
    	> & 2 + p_t\varepsilon - (1 + \varepsilon + p_t\varepsilon) \\
    	=& f(\{s_{\rm big}\}, x^1_{-j_1}) - f(x^1_{-j_1})  \\
    	=& U_{j_1}(\{s_{\rm big}\}, x^1_{-j_1})
    \end{align}
    For agent $j_t, t>1$, the action set is $\{\{s_{\rm sm}\}, \{s_t\}\}$, implying that
    \begin{align}
    	U_{j_t}(x^t) =& 1 + \varepsilon \\
    	> & 1 + p_t\varepsilon -  p_t\varepsilon \\
    	=& f(\{s_t\}, x^t_{-j_t}) - f(x^t_{-j_t})  \\
    	=& U_{j_t}(\{s_t\}, x^t_{-j_t})
    \end{align}
    Since all other agents have only a single action in their action sets, we conclude that $x^\eq$ is an equilibrium action profile. The optimal profile $x^\opt$ is where $j_1$ chooses $\{s_{\rm big}\}$, $j_t$ chooses $\{s_t\}$ for $t>1$, and all other agents choose $\{s_\varepsilon\}$, implying that $f(x^\opt) = 2 + \tau(G)-1 + \varepsilon$. Therefore,
    \begin{equation}
    	\frac{f(x^\eq)}{f(x^\opt)} = \frac{1 + 2\varepsilon}{1 + \tau(G) + \varepsilon}.
    \end{equation}
    As $\varepsilon \to 0$, we see that \eqref{eq:vug_poa_ub} holds.
\end{proof}

\section{Consistent Valid Utility Games} \label{sec:graph_rev}

The previous section showed that the performance guarantees associated with valid utility games can be quite poor, even for information sharing constraint graphs that are quite dense. It is thus imperative to find other classes of utility functions that would offer more robustness against such constraints. To this end, we introduce an additional utility function property called \emph{consistency}. A utility function is consistent if
\begin{equation} \label{eq:cons}
	U_i(x_i, x_A) \ge U_i(x_i, x_B),
\end{equation}
for all $A \subseteq B \subseteq N \setminus \{i\}$, for all $x_i \in X_i, x_A \in \Pi_{j \in A} x_i, x_B \in \Pi_{j \in B} x_j$,  and for all $i \in N$. Here the sets $A$ and $B$ represent possible choices of incoming neighbors; the consistency property simply states that an agent's preference for any action decreases as the set of agents that it can observe grows. Many common choices of utility functions, including the marginal contribution utility function in \eqref{eq:mcdef}, satisfy this property.

\begin{figure}
	\subfloat[In this graph, there are 4 cliques of size 1 (one for each node), 5 cliques of size 2 (one representing each edge), and 2 cliques of size 3 (the sets $\{1,2,3\}$ and $\{1, 2, 4\}$). 
	The maximum independent set is $\{3, 4\}$, thus $\alpha(G)=2$. For this graph, $\alpha^*(G)=\alpha(G)$.\label{fig:graph}]{%
		\makebox[0.48\textwidth]{\includegraphics[scale=0.4]{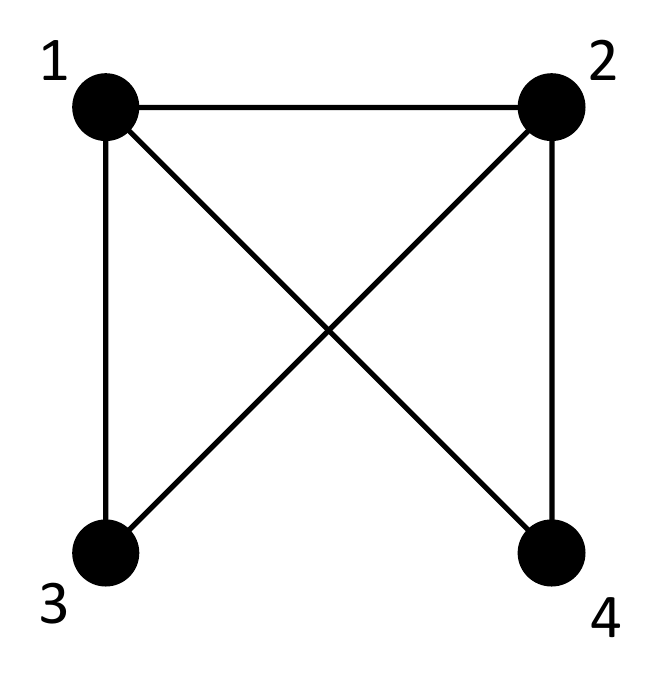}}
}

	\subfloat[A graph where $\alpha(G) = 2$ and $\alpha^*(G) = 2.5$, where $z = {[1/2, 1/2, 1/2, 1/2, 1/2]}^T$ maximizes \eqref{eq:fracindlp}. As a note, this is the graph with the fewest number of nodes and edges such that $\alpha(G) < \alpha^*(G)$. \label{fig:frac_ind}  ]{%
		\makebox[0.48\textwidth]{\includegraphics[scale=0.4]{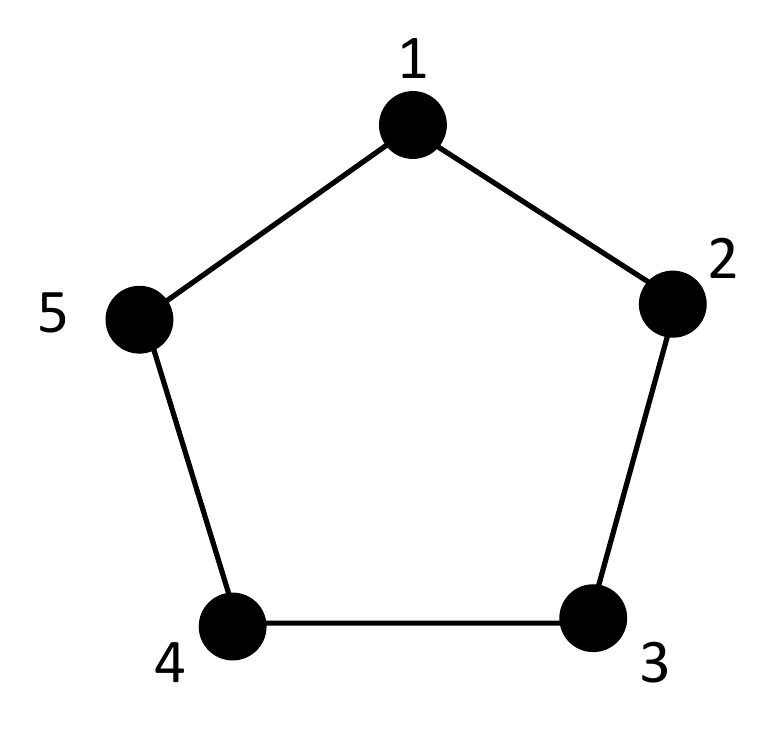}}
	}
	\caption{Two example graphs showcasing the graph properties defined in Section \ref{sec:graph_rev}.}
	\label{fig:ex}
\end{figure}

In order to state the result for this section, we first introduce some terms from graph theory. We begin with cliques:\footnote{The terms \emph{clique} and \emph{independence set} are traditionally defined only for undirected graphs, however, we adapt those terms for our purposes here.} a \emph{clique} is a set of nodes $C \subseteq V$ such that for every $i, j \in C$, either $(i, j) \in E$ or $(j, i) \in E$.  We denote by $K(G)$ the set of all cliques in $G$. 

Another important notion in graph theory is that of independence.  An \emph{independent set} $J \subseteq V$ is a set of vertices such that $v_1, v_2 \in J$ implies $(v_1, v_2), (v_2, v_1) \notin E$. A \emph{maximum independent set} is an independent set of $G$ such that no other independent set has more vertices. The \emph{independence number} $\alpha(G)$ is the number of nodes in the largest independent set in $G$. For an example, see Figure \ref{fig:graph}. 

The work in \cite{Godsil2001a} equivalently characterizes the independence number as the solution to an integer linear program~\footnote{It is actually the chromatic number and clique number that are defined this way in \cite{Godsil2001a}. However, using graph complementarity, it is an easy extension to show that the solution to the linear program in \eqref{eq:indlp} yields a maximum independent set.}. Let $Q \in \mathbb{R}^{|K(G)| \times n}$ be the binary matrix whose rows are indicator vectors for the cliques in $G$. In other words, $Q_{ij} = 1$ if node $j$ belongs to clique $i$ in $G$, and 0 otherwise. Note that $Q$ also includes cliques of size 1 (the individual nodes). Then $\alpha(G)$ is given by
\begin{equation}
	\label{eq:indlp}
	\begin{aligned}
		& \max_z
		& & z^T \textbf{1} \\
		& \text{subject to}
		& & Qz \leq \textbf{1} \\
		& & & z \in \mathbb{Z}^n \geq \textbf{0}.
	\end{aligned}
\end{equation}

Note by definition that $\alpha(G)$ is always a positive integer. However, in many applications, it is helpful to consider a real-valued relaxation on these notions: this is the motivation for fractional graph theory \cite{Godsil2001a}. Here we leverage the \emph{fractional independence number} $\alpha^*(G)$, which we define as the real-valued relaxation to \eqref{eq:indlp}: \footnote{Another defintion of fractional independence exists in the literature (see \cite{Arumugam2007a}), which was created to preserve certain properties of graph independence (such as nested maximality), but has not been shown to preserve $\alpha^*(G) = \omega^*(\bar{G})$, where $\bar{G}$ is the complement graph of $G$ and $\omega^*(G)$ is the fractional clique number of $G$.}

\begin{equation}
	\label{eq:fracindlp}
	\begin{aligned}
		\alpha^*(G) := & \max_z
		& & z^T \textbf{1} \\
		& \text{subject to}
		& & Qz \leq \textbf{1} \\
		& & & z \geq \textbf{0}.
	\end{aligned}
\end{equation}  

Considering now valid utility functions that are also consistent, we have the following result.

\begin{theorem} \label{thm:cons}
	For any graph $G$,
	\begin{equation}
		\poa(\mathcal{H}_c, G) \ge \frac{1}{1 + \alpha^*(\bar{G})},
	\end{equation}
	where $\mathcal{H}_c \subseteq \mathcal{H}$ is the set of all consistent valid utility games, and $\bar{G}$ is the subgraph of $G$ such that any ``non-reciprocal" edges from $G$ are removed, i.e. if $\bar{G} = (\bar{V}, \bar{E})$, then $\bar{V} = V$, and $(i, j) \in \bar{E}$ iff $(i, j), (j, i) \in E$.
\end{theorem}

The proof is given in Appendix~\ref{app:cons}. The consistency property allows one to make a stronger guarantee about the set of resulting equilibria. For instance, consider again the example in Figure~\ref{fig:info_grp}. Of course, the complete graph on the left is such that $\alpha^*(\bar{G}) = \alpha^*(G) = \alpha(G) = 1$. Therefore, Theorem~\ref{thm:cons} gives the same bound as the more general case: that $\poa(\mathcal{H}_c, G) \ge 1/2$. The rightmost graph $G$ is such that $\bar{G}$ is a line graph: edges (2, 1), (1, 3), and (3, 4) are removed since they have no reciprocal. Here $\alpha^*(\bar{G}) = \alpha(\bar(G)) = 2$, ensuring that $\poa(\mathcal{H}_c, G)\ge 1/3$, compared to $1/5$ for the more general case. In fact, it is trivial to show that $\tau(G) \ge \alpha(\bar{G})$ for any $G$, since if $2$ nodes are in the same information group in $G$, they cannot be independent in $\bar{G}$. Therefore, one is better off (in terms of equilibrium guarantees) implementing consistent utilities within the valid utility framework.

\section{A Bound on Optimal Utilities}

In this section, we relax the assumption that the system is a valid utility game. Instead, we consider the class of all utility functions, and we show an upper bound on the price of anarchy given the information sharing constraint graph $G$.

\begin{figure}
	\centering
	\includegraphics[scale=0.5]{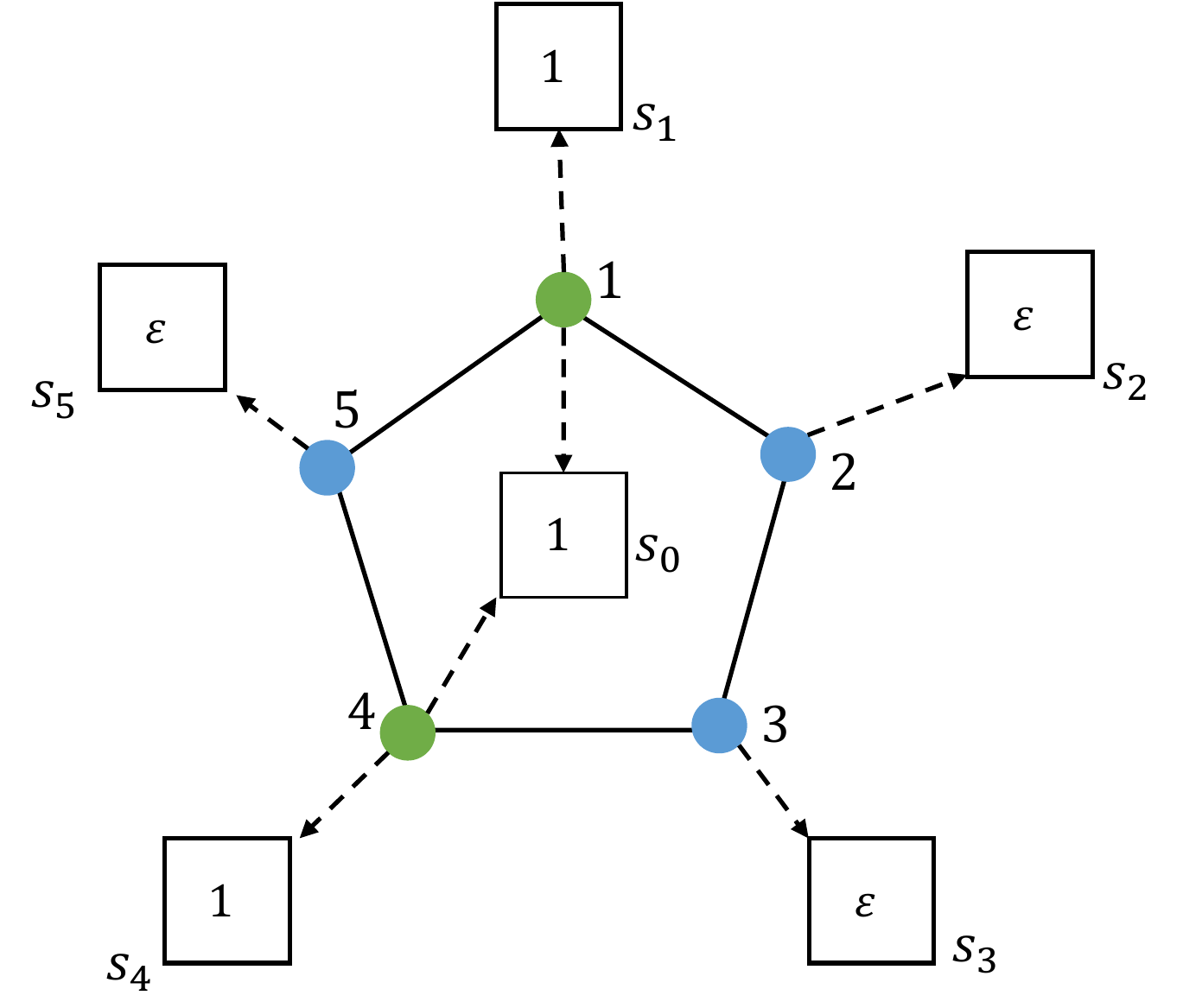}
	\caption{An example for the proof of Proposition \ref{prop:svug_ub}. The agents are labeled $\{1, \dots, 5\}$, the solid lines represent the graph $G$, and the dashed lines represent the actions available to each agent. $G$ is a ring graph with $5$ agents, and $J = \{1, 4\}$ (the green nodes). In a worst-case equilibrium, the green agents choose $s_0$ and the rest choose $s_i$. The optimal choices are for the green agents to choose $s_i$ and one of the blue agents choose $s_0$.}
	\label{fig:svug_ub}
\end{figure}

\begin{prop}\label{prop:svug_ub}
	For any admissible utility function profile $U = (U_1, \dots, U_n)$ and any graph $G$,
	\begin{equation}
		\poa(\mathcal{H}_U(G)) \le \frac{1}{\alpha(G)},
	\end{equation}
	where $\mathcal{H}_U(G) \subseteq \mathcal{H}$ is the set of systems that employ $U$ and are subject to the information sharing constraint graph $G$. 
\end{prop}

For many graphs, there is still a large gap between the upper bound on $\poa$ shown in Proposition~\ref{prop:svug_ub} and the lower bound for consistent valid utilities shown in Theorem~\ref{thm:cons}. For instance, if $G$ is a fully-connected directed acyclic graph, then $\poa(\mathcal{H}_c(G)) \ge 1/(n+1)$. However, it has been shown in \cite{Gharesifard2017} that deploying the marginal contribution utility can guarantee a $\poa$ of $1/2$ for this graph constraint. Thus utilities which are optimal in this sense are an ongoing study of future work. 

\begin{proof}
	This proof is given by example. For ease of notation, denote $\alpha$ to mean $\alpha(G)$ and let $J \subseteq N$ to be a fixed maximum independent set. Consider a system with base set of  resources $S = \{s_0, \dots, s_n\}$. Let $f(s_0) = 1$, let $f(s_i) = 1$ for $i \in J$, and let $f(s_i) = \varepsilon$ for $i \notin J$ and for some small $\varepsilon$. For every agent $i \in J$, the action set is $X_i = \{\{s_0\}, \{s_i\}\}$. For every agent $i \notin J$, the action set is $X_i =\{\{s_i\}\}$, in other words these agents have only a single action to choose. See Figure~\ref{fig:svug_ub} for an example.
	
	Based on $G$ agents in $J$ cannot have a utility which directly accounts for the action of any other agent in $J$ at equilibrium. One can assume without loss of generality that for $i \in J$, $U_i(s_i) \le U_i(s_0)$, since the two elements are indistinguishable except by indexing, which could easily be switched. Therefore, a worst-case equilibrium decision set $x^\eq$ would be all agents in $J$ choose $\{s_0\}$ and all other agents choose $\{s_i\}$. In this case $f(x^\eq) = 1 + (N - \alpha)\varepsilon$. On the other hand, the optimal action profile $x^\opt$ is where all agents choose $s_i$, implying that $f(x^\opt)= \alpha + 1 + \varepsilon + (N - \alpha - 1)\varepsilon$. Then
	\begin{equation}
		\lim_{\varepsilon \to 0} \frac{f(x^\eq)}{f(x^{\rm opt})} = \lim_{\varepsilon \to 0} \frac{1 + (N - \alpha)\varepsilon}{ \alpha  + (N - \alpha)\varepsilon} = \frac{1}{\alpha}.
	\end{equation}
	By definition, this is then an upper bound on $\poa$.
\end{proof}

\section{Conclusion}

In this paper we have explored how information sharing constraints can affect the value of the resulting Nash equilibria in valid utility games. Specifically, we showed that the performance guarantees degrade quickly as information sharing constraints are imposed. In order to mitigate these effects, we introduce the notion of a consistent utility function, and show that the performance guarantees are often strictly better when restricting to this set of utilities. Finally, we gave an upper bound on performance guarantees for any set of utilities.

Future work will continue to understand utility design that is robust against these information sharing constraints. It will also focus applying this analysis to other types of information sharing constraints: for instance, a constrained number of bits or the action of only one agent. 

\bibliographystyle{ieeetr}
\bibliography{references.bib}

\appendix

\subsection{Proof for Theorem~\ref{thm:cons}} \label{app:cons}

This proof requires the introduction of a few more graph terms. A \emph{clique cover} is a partition on $V$ such that the nodes in each set of the partition form a clique. The \emph{clique cover number} $k(G)$ is the minimum number of sets needed to form a clique cover of $G$. For example in Figure \ref{fig:graph}, a minimum clique cover is $\{1, 3\}, \{2, 4\}$, so $k(G)=2$.

Similar to the independence number, it can be shown that $k(G)$ is equivalently defined as an integer linear programming problem. In fact this ILP is the dual to \eqref{eq:indlp}, implying that $\alpha(G) \leq k(G)$. Likewise, $k^*(G)$, the \emph{fractional clique cover number} of $G$, can be defined by the dual to \eqref{eq:fracindlp}:

\begin{equation}
	\label{eq:fraccliquecovlp}
	\begin{aligned}
		k^*(G) := & \min_y
		& & y^T \textbf{1} \\
		& \text{subject to}
		& & Q^Ty \geq \textbf{1} \\
		& & & y \geq \textbf{0}.
	\end{aligned}
\end{equation}
In accordance with the Strong Duality of Linear Programming \cite{matousek2007understanding}, it follows that:
\begin{equation}
	\label{eq:fracprop}
	\alpha(G) \leq \alpha^*(G) = k^*(G) \leq k(G).
\end{equation}
An example of a graph where the independence number differs from the fractional independence number is found in Figure~\ref{fig:frac_ind}.

Begin with
	\begin{align}
		f(x^\opt) \le & f(x^\eq) + \Delta(x^\opt | x^\eq)  \label{eq:cons_lb1},\\
		= & f(x^\eq) + \sum_i \Delta(x^\opt_i | x^\opt_{1:i-1}, x^\eq) \label{eq:cons_lb2},\\
		\le & f(x^\eq) + \sum_i \Delta(x^\opt_i | x^\eq_\nei) \label{eq:cons_lb3}, \\
		\le & f(x^\eq) + \sum_i U_i(x^\opt_i, x^\eq_\nei) \label{eq:cons_lb4}, \\
		\le & f(x^\eq) + \sum_i U_i(x^\eq_i, x^\eq_\nei) \label{eq:cons_lb5},
	\end{align}
	where \eqref{eq:cons_lb1} and \eqref{eq:cons_lb3} are true by submodularity of $f$, \eqref{eq:cons_lb4} is true from \ref{itm:marg}) of Definition~\ref{def:vug}, and \eqref{eq:cons_lb5} is true by definition of equilibrium. Now suppose that we have a set of scalars $\{y_k\}_{k \in K(\bar{G})}$, such that $y_k \ge 0$ and $\sum_{k: i\in k} y_k \ge 1$ for all $i$. Then
	\begin{align}
		\sum_i U_i(x_i^\eq, x^\eq_\nei) \le & \sum_i U_i(x_i^\eq, x^\eq_\nei) \left(\sum_{k \in K(\bar{G}): i \in k} y_k \right)  \\
		= & \sum_i \sum_{k \in K(\bar{G}): i \in k} y_k U_i(x_i^\eq, x^\eq_\nei) \\
		= & \sum_{k \in K(\bar{G})} \sum_{i \in k}  y_k U_i(x_i^\eq, x^\eq_\nei) \\
		\le & \sum_{k \in K(\bar{G})} \sum_{i \in k}  y_k U_i(x_i^\eq, x^\eq_{k \setminus \{i\}}) \label{eq:cons_lb6} \\
		\le & \sum_{k \in K(\bar{G})} y_k \sum_{i \in k} U_i(x_i^\eq, x^\eq_{k \setminus \{i\}})  \label{eq:cons_lb7}\\
		\le & \sum_{k \in K(\bar{G})} y_k f(x^\eq_k) \label{eq:cons_lb8} \\
		\le & f(x^\eq) \sum_{k \in K(\bar{G})} y_k,
	\end{align}
	where \eqref{eq:cons_lb6} is true by the consistency property, \eqref{eq:cons_lb7} is true by \ref{itm:sum}) of Definition~\ref{def:vug}, and \eqref{eq:cons_lb8} is true by the monotonicity of $f$. Combining this with \eqref{eq:cons_lb5} yields
	\begin{equation}
		\frac{f(x^\eq)}{f(x^\opt) }\ge \frac{1}{1 + \sum_{k \in K(\bar{G})}y_k}.
	\end{equation}
	The choice of $\{y_k\}_{k \in K(\bar{G})}$  that minimizes $\sum_{k \in K(\bar{G})}y_k$ will therefore give the highest lower bound. One can formulate this as
	\begin{equation}
		\begin{array}{cc}
			\min_{\{y_k\}_{k \in K(\bar{G})}} &  \sum_{k \in K(\bar{G})}y_k \\
			\text{subject to } & \sum_{k: i \in k} y_k \ge 1, \text{ for all } i \\
			& y_k \ge 0, \text{ for all } k.
		\end{array}
	\end{equation}
	This is equivalent to the formulation of $k^*(\bar{G})$ in \eqref{eq:fraccliquecovlp}. Since $k^*(\bar{G}) = \alpha^*(\bar{G})$, this completes the proof.

\end{document}